\documentclass[12pt,a4paper]{article}
\usepackage{epsf}
\usepackage{latexsym}

\newcommand{\xbj}{x_{B}}              
\newcommand{\xn}{x_{N}}               

\newcommand{\eqpt}{\hspace{6pt}.\hspace{6pt}}  
\newcommand{\eqcm}{\hspace{6pt},\hspace{6pt}}

\newcommand{\GeV}{\mbox{\ GeV}}

\newcommand{\re}{\mbox{Re}\,}

\newcommand{\mat}[3]{\langle #1 \, | \, #2 \, | \, #3 \rangle}

\renewcommand{\in}{\raise -4pt\hbox{\tiny in}}
\newcommand{\out}{\raise -4pt\hbox{\tiny out}}

\newcommand{\Journal}[4]{#1 #2 (#4) #3}
\newcommand{\NPB}{Nucl.\ Phys.\ B}
\newcommand{\PLB}{Phys.\ Lett.\ B}
\newcommand{\PR}{Phys.\ Rep.}
\newcommand{\PRL}{Phys.\ Rev.\ Lett.}
\newcommand{\PRD}{Phys.\ Rev.\ D}
\newcommand{\ZPC}{Z.\ Phys.\ C}

\begin{document}

\begin{flushright}
SPhN--98--01\\
CPT--S593--1297
\end{flushright}
\vspace{2\baselineskip}

\begin{center}
{\Large\bf TIME ORDERING IN \\ \medskip
           OFF-DIAGONAL PARTON DISTRIBUTIONS}\\
\vspace{\baselineskip}
{Markus Diehl}\\ \medskip
{\small\it CPT, Ecole Polytechnique, 91128 Palaiseau,
France\\[0.1\baselineskip]
present address:\\
DAPNIA/SPhN, C. E. Saclay, 91191 Gif sur Yvette, France}\\
\vspace{\baselineskip}
and\\
\vspace{\baselineskip}
{Thierry Gousset}\\ \medskip
{\small\it NIKHEF, P. O. Box 41882, 1009 DB Amsterdam, The
Netherlands\\[0.1\baselineskip]
present address:\\
SUBATECH, B. P. 20722, 44307 Nantes CEDEX 3, France}\\
\vspace{2\baselineskip}
{\bf Abstract}\\[\baselineskip]
\parbox{0.9\textwidth}{We investigate the relevance of time ordering
  in the definition of off-diagonal parton distributions in terms of
  products of fields. The method we use easily allows determination of
  their support properties and provides a link to their interpretation
  from a parton point of view. It can also readily be applied to meson
  distribution amplitudes.}
\end{center}
\vspace{\baselineskip}

\begin{figure}
  \epsfysize 3cm
  $$\epsfbox{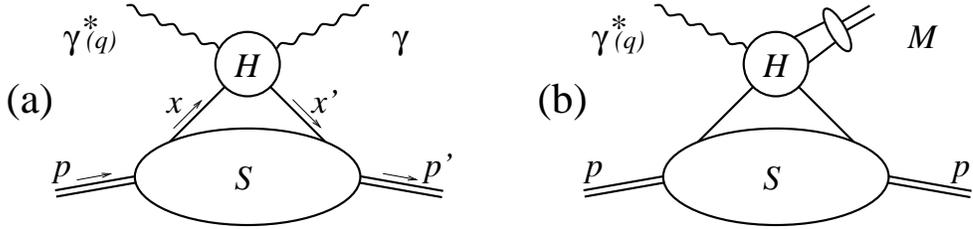}$$
  \caption{\label{fig:factorise}Diagrams for $\gamma^\ast p \to A p$
    at large $Q^2$ and small $t$, where $A$ is $(a)$ a real photon or
    $(b)$ a meson $M$. They consist of an off-diagonal parton
    distribution, a hard scattering part, and a meson distribution
    amplitude in case $(b)$. The partons connecting $S$ and $H$ have
    momenta $k$ and $k'$ with plus components $k^+ = x p^+$ and $k'^+
    = x' p^+$.}
\end{figure}

\noindent
1. Recently there has been renewed interest in off-diagonal parton
distributions\footnote{The terms nondiagonal, off-forward or
nonforward parton distributions are also used in the literature.},
i.e.\ in correlation functions of quark or gluon fields between
different nucleon states, which have been introduced some time ago
\cite{oldies}. They appear for instance in the description of
exclusive photon or meson production in $\gamma^\ast p$ collisions at
large photon virtuality $Q^2$ and small squared momentum transfer $t$
of the proton~[2--5], showing up as the long distance nonperturbative
quantities when one factorises the transition amplitude into short and
long distance subprocesses.\footnote{The question of factorisation has
been discussed in~\cite{Col} for the production of mesons and
in~\cite{RadLong} for the production of a real photon.} In meson
electroproduction there is yet another nonperturbative ingredient,
namely the meson distribution amplitude which describes the transition
from the valence quarks to the final vector meson. The corresponding
diagrams are shown in Fig.~\ref{fig:factorise}.

The relevance of off-diagonal parton distributions is not restricted
to these processes. In particular the asymmetric gluon distribution
has been considered in several $\gamma^\ast p$ processes at small
Bjorken $x$, such as diffractive production of dijets and of heavy
quarks \cite{Diffraction}. For exclusive $J/ \psi$ production one may
relax the requirement of large $Q^2$ because of the large meson mass
\cite{Jpsi}, as well as for the exclusive production of a $Z$ boson
\cite{Zboson}.

\vspace{\baselineskip}
\noindent
2. Let us first have a closer look at the kinematics of the reactions 
we have in mind. We denote particle momenta according to
\begin{equation}
  \label{reaction}
  \gamma^\ast(q) + p(p)\to A(q') + p(p')  \eqcm
\end{equation}
where $A$ is a meson, real photon or heavy gauge boson, or a
quark-antiquark pair in a diffractive process. One might also have a
virtual $Z$ or $W$ instead of the $\gamma^\ast$ and/or $A$ may be
charged, replacing the initial or final state proton by a neutron as
necessary. We will use the momentum transfer $\Delta = p - p'$ from
the proton, and the invariants $t = \Delta^2$, $W^2 = (p + q)^2$, $Q^2
= - q^2$, $m_A^2 = q'^2$, $m_p^2 = p^2$.

We now consider those reference frames where the incoming photon and
proton are collinear. Picking out any one of them we choose the $z$
axis along the proton momentum $p$ and introduce two lightlike vectors
\begin{equation}
  \label{vectors}
  v  = \frac{1}{\sqrt{2}} \, (1, 0, 0, 1)  \eqcm \hspace{2em}
  v' = \frac{1}{\sqrt{2}} \, (1, 0, 0, -1) \eqcm 
\end{equation}
which respectively set a plus and minus direction. We can write
\begin{equation}
  \label{proton-kinematics}
  p^\mu = p^+ \, v^\mu + \frac{m_p^2}{2 p^+} \, v'^\mu  \eqcm
  \hspace{2em}
  \Delta^\mu = \xi p^+ \, v^\mu - \frac{m_p^2 \xi + {\bf
  \Delta}_T^2}{2 p^+ (1 - \xi)} \, v'^\mu + \Delta_T^\mu
\end{equation}
and
\begin{equation}
  \label{kinematics-t}
  t = - \frac{m_p^2 \xi^2 + {\bf \Delta}_T^2}{1 - \xi}  \eqcm
\end{equation}
having defined the variable $\xi = \Delta^+ / p^+$, whose exact
expression in terms of the invariants listed above we will not need
here. We further have
\begin{equation}
  \label{photon-kinematics}
  q^\mu = - \xn p^+ \, v^\mu + \frac{Q^2}{2 p^+ \xn} \, v'^\mu  \eqcm
\end{equation}
where
\begin{equation}
  \label{Nachtmann}
  \xn = \frac{2 \xbj}{1 + \sqrt{1 + 4 \xbj^2 m_p^2 /Q^2}}
\end{equation}
is Nachtmann's and $\xbj = Q^2 / (2 p \cdot q)$ Bjorken's scaling
variable. From the positivity of particle energies in the final state
one obtains
\begin{equation}
  x_N \le \xi < 1  \eqpt
\end{equation}

To ensure that the blob $H$ in Fig.~\ref{fig:factorise} corresponds to
a hard scattering we require that at least one of $Q^2$ and $m_A^2$ be
large compared with a GeV$^2$. For the blob $S$ to be a soft, long
distance quantity one will need that the transverse bend $|
{\bf\Delta}_T |$ is of the order of a hadronic scale.

The threshold region is special in its dynamics: apart from the
formation of resonances one can expect strong rescattering effects
between $A$ and the outgoing proton which may destroy
factorisation. One will ask for $W - m_A - m_p \gg 1 \GeV$ to exclude
this region.\footnote{We remark that we will not actually make use of
this condition in the arguments of this paper.} It can be shown that
this is equivalent to limiting the range of $\xi$ to $1 - \xi \gg
(m_p\, W) /(W^2 + Q^2)$, which according to (\ref{kinematics-t}) also
puts an upper limit on $-t$. Up to corrections of order $m_p^2 / (W^2
- m_A^2)$ and ${\bf \Delta}_T^2 / (W^2 - m_A^2)$ one then has
\begin{equation}
  \label{xi}
  \xi = \frac{Q^2 + m_A^2}{2 p \cdot q}  \eqpt
\end{equation}

Under these conditions one can find a frame, e.g.\ the $\gamma^\ast p$
c.m., where the initial proton is moving fast and collides head on
with the photon, $p^+ \gg m_p$, and where the scattered proton is fast
as well, $p'^+ \gg m_p$. Such a frame is natural for a partonic
interpretation of our process.\footnote{The analysis we perform in
this paper may also be done in frames where $p$ and $q$ have nonzero
transverse components~\cite{Ji}, provided that these are sufficiently
small, as ${\bf \Delta}_T$ in the frames considered here.}

\vspace{\baselineskip}
\noindent
3. The nonperturbative transition from the proton to the parton level
(see Fig.~\ref{fig:factorise}) is described by a two-point function of
the form
\begin{equation}
  \label{distribution}
  M = \frac{1}{2 \pi}  \int dz^- \,
  e^{i x p^+ z^-} \left. \mat{p', \sigma'}{T \, \bar{\psi}(0) \gamma^+
  \psi(z)}{p, \sigma} \right|_{z = z^- \, v'}  \eqpt
\end{equation}
This representation holds in the $A^+ = 0$ gauge, where $A^\mu$ is the
gluon potential; for other gauges one has to insert the standard path
ordered exponential between the two quark fields. Instead of
$\gamma^+$ in (\ref{distribution}) there can be other Dirac matrices,
which in order to give a leading twist contribution must transform
like $\gamma^+$ under a longitudinal boost. Which matrices are
relevant depends on the process considered; in virtual Compton
scattering, $\gamma^\ast p \to \gamma p$, one needs for instance
$\gamma^+$ and $\gamma^+ \gamma_5$. There are also contributions from
two-point functions where gluon field strengths replace the quark
fields in (\ref{distribution}).

We notice that these quark and gluon two-point functions depend on the
spins $\sigma$ and $\sigma'$ of the initial and final state
proton. This spin structure can be made explicit by writing
(\ref{distribution}) as a sum of Dirac bilinears for the proton
multiplied by scalar functions $F_i(x,\xi,t)$. It is these functions
that are called off-diagonal parton distributions; we will not need
their explicit definitions here.\footnote{A comparison of the various
distributions introduced in the recent literature can be found in
\cite{RadLong}.} For ease of writing we will not explicitly display
the labels $\sigma$ and $\sigma'$ henceforth.

The usual parton distributions that occur in inclusive deep inelastic
scattering involve diagonal matrix elements of quark fields or gluon
field strengths, i.e.\ they correspond to the limit $p' = p$ and
$\sigma' = \sigma$ in (\ref{distribution}). The variables $\xi =
\Delta^+ / p^+$ and $t = \Delta^2$ are then zero since $\Delta =
0$. The diagrams that describe inclusive deep inelastic scattering are
obtained by cutting the ones of Fig.~\ref{fig:factorise} $(a)$ with
the real photon replaced by a virtual one.

At this point we note that because the proton states in
(\ref{distribution}) are different the off-diagonal distributions do
not have any probabilistic interpretation, contrary to the usual
parton distributions. In particular one cannot expect the $F_i$ to
have a definite sign as ordinary quark or gluon densities, which are
nonnegative for $0<x<1$.

\vspace{\baselineskip}
\noindent
4. When calculating the transition amplitude from the diagrams in
Fig.~\ref{fig:factorise} the soft blob describing the proton-parton
coupling translates into the Fourier transformed matrix element of a
\emph{time ordered} product of parton operators. We will show that one
can actually drop the time ordering and instead use \emph{ordinary}
products.

The same problem has long ago been considered by Jaffe \cite{Jaf} for
the parton distributions in deep inelastic scattering. For the deep
inelastic cross section one needs the \emph{absorptive part} of the
forward $\gamma^\ast p \to \gamma^\ast p$ amplitude, cutting the
diagrams including the soft blob, and naturally obtains ordinary
products instead of time ordered ones. It was however shown in
\cite{Jaf} that the ordinary product already appears in the
\emph{full} amplitude, not only its absorptive part.\footnote{This is
also plausible from the point of view of the operator product
expansion. There one obtains matrix elements of local operators, which
are moments of the parton distributions; only the Wilson coefficients
describing the hard scattering depend on whether one takes the full
amplitude or its imaginary part.} The same result also follows from
early work by Landshoff and Polkinghorne \cite{Lan}, without being
explicitly stated there, and we will apply their method to the
non-diagonal case here.

The fact that one can omit time ordering in the parton distributions
has important consequences for their support properties in the scaling
variable $x$, which in turn are crucial for their interpretation in
terms of partons~\cite{Jaf,EllFurPet}. In the off-diagonal case it is
also needed if one wants to derive a dispersion relation for the
scattering amplitude~\cite{Col}. Furthermore it allows one to constrain
the two-point function (\ref{distribution}) using time reversal
invariance: together with parity invariance one finds that the phases
of the scalar parton distributions $F_i(x,\xi,t)$ are a matter of 
convention, so that they can be defined as real valued.

As a by-product of the demonstration we will obtain the support
properties for $F_i$ in $x$ and identify different regions in $x$
according to whether the partons are in the initial or final state of
the hard scattering $H$.\footnote{The support of $F_i$ has been
obtained in~\cite{RadLong} by analysing the soft blob $S$ in terms of
Feynman graphs.} This generalises the results obtained in~\cite{Jaf}
concerning the support and partonic interpretation of the usual
densities to the case of off-diagonal distributions.

\vspace{\baselineskip}
\noindent
5. Let us first consider the quark two-point function $M$ of
(\ref{distribution}). Our arguments will be unchanged if $\gamma^+$ is
replaced by another suitable Dirac matrix. We rewrite
\begin{equation}
  \label{rewrite}
  M = \frac{1}{(2\pi)^4}  \int d^2 {\bf k}_T \, d k^- \, 
  \left. {\cal A} \right|_{k^+ =x p^+}
\end{equation}
where
\begin{equation}
  \label{amplitude}
  {\cal A} = \int d^4 z \, e^{i k \cdot z} \mat{p'}{T \, \bar{\psi}(0)
  \gamma^+ \psi(z)}{p}  
\end{equation}
is an amplitude describing the scattering of an off-shell antiquark
with momentum $-k$ on an on-shell proton with momentum $p$,
\begin{equation}
  \label{scattering}
  p(p) + \bar{q}(-k) \to p(p') + \bar{q}(-k')  \eqpt
\end{equation}
Note that this amplitude is not truncated in the parton legs. The plus
components $k^+ = x p^+$ and $k'^+ = x' p^+$ with $x' = x - \xi$ are
kept fixed in the following.

The kinematical invariants $\cal A$ depends on are the virtualities
$k^2$ and $k'^2$ and the Mandelstam variables $s = (-k + p)^2$, $u =
(k + p')^2$ and $t$, the latter being fixed by the kinematics of the
reaction (\ref{reaction}). Following~\cite{Lan} we assume that the
analytical properties of $\cal A$ are the usual ones known from the
study of Feynman diagrams, i.e.\ that it has cuts for nonnegative $\re
s$, $\re u$ and singularities for nonnegative $\re k^2$ and $\re
k'^2$. In the standard conventions, these singularities are slightly
below the real axis of these variables.\footnote{To avoid possible
complications due to the dependence of ${\cal A}$ on the spin of the
proton one can apply our argument directly to the scalar functions
$F_i(x,\xi,t)$ where the proton spin structure has been factored
off. The corresponding amplitudes for (\protect\ref{scattering}) only
depend on the above invariants and on the plus components $p^+$,
$\Delta^+$ and $k^+$ which are all fixed. The dependence on plus
components comes from factoring off the proton spin and from the Dirac
matrix between the quark fields, and also from the choice of gauge
$A^+ = 0$.}

We can carry out the integration over $k^-$ in (\ref{rewrite}) taking
into account the singularity structure of $\cal A$. To see where the
singularities are located in the complex $k^-$ plane we in turn
express $k^-$ through one of the invariants $s$, $k'^2$, $k^2$, $u$,
supplemented by $x$, ${\bf k}_T$ and the variables in
(\ref{proton-kinematics}) which are all kept fixed during the $k^-$
integration:
\begin{eqnarray}
  \label{parton-kinematics}
k^- &=& {s + {\bf k}_T^2 \over 2 p^+ (x-1)} + p^- \eqcm \\ \nonumber
k^- &=& {k'^2 + ({\bf k}_T - {\bf\Delta}_T)^2 \over 2 p^+ (x-\xi)} +
\Delta^-  \eqcm \\ \nonumber
k^- &=& {k^2 + {\bf k}_T^2 \over 2 p^+ x}  \eqcm \\ \nonumber
k^- &=& {u + ({\bf k}_T - {\bf\Delta}_T)^2 \over 2 p^+ (x+1-\xi)} -
p^- + \Delta^- \eqpt
\end{eqnarray}
There is a sequence of sign changes of the above denominators as $x$
varies from $-\infty$ to $+\infty$. Using $0 \le \xi < 1$ we see that
they occur at $x = \xi-1$, $0$, $\xi$ and $1$. These signs determine
whether the singularities of $\cal A$ are located above or below the
real axis in the $k^-$ plane.

For $x \le \xi-1$ and $x \ge 1$ all singularities lie on the same side
of the real $k^-$ axis and we can close the contour in the other half
plane without encircling any singularity. The $k^-$ integral vanishes
and we deduce that
\begin{quote}
  \centering $M$ is only nonzero for $\xi-1 < x < 1$.
\end{quote}
To be able to close the integration contour in the $k^-$ plane we have
made the dynamical assumption that $\cal A$ vanishes sufficiently fast
as $k^-$ tends to infinity in the complex plane while $k^+$ and ${\bf
k}_T$ are fixed, a limit in which the parton virtualities $k^2$ and
$k'^2$ become infinite. Note that we have not yet integrated over
${\bf k}_T \,$: as remarked in~\cite{Jaf} the integration of $\cal A$
over ${\bf k}_T$ and $k^-$ leads to ultraviolet divergences, but this
does not contradict our assumption that the integral over $k^-$ alone
is sufficiently well behaved. The full integration over ${\bf k}_T$
and $k^-$ can only be done in a suitably regularised theory; it is
known from the usual parton distributions that the corresponding
renormalisation of the integral (\ref{rewrite}) is intimately related
with their logarithmic evolution in QCD.

\begin{figure}[t]
  \epsfysize 10cm
  $$\epsfbox{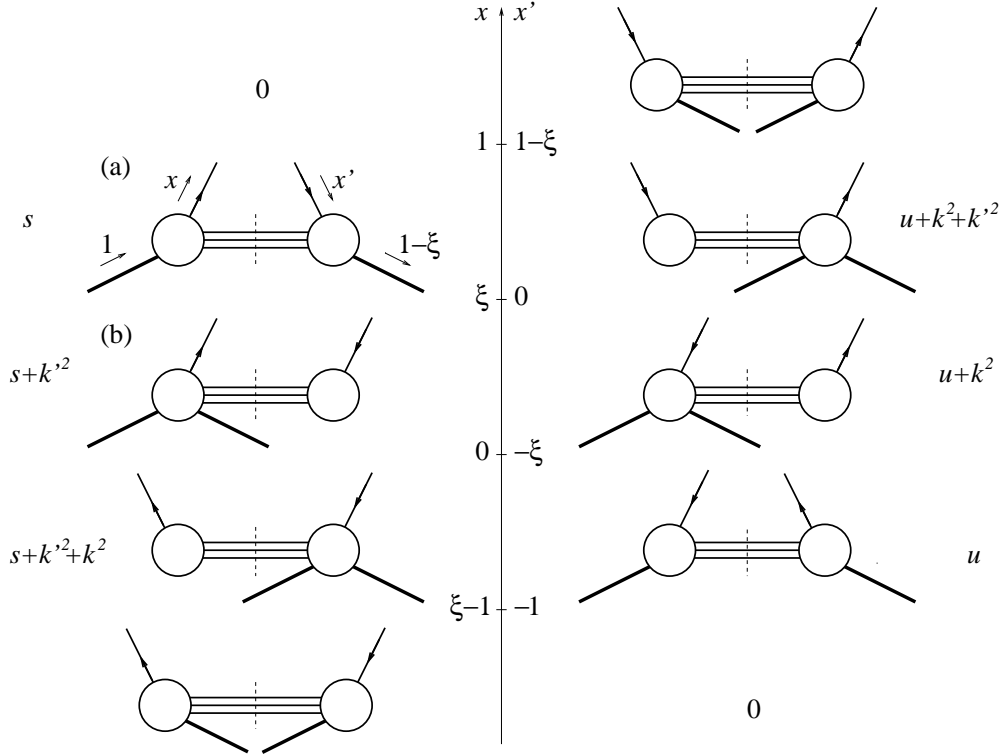}$$ 
\caption{\label{fig:cuts}The two alternatives to pick up singularities
  in the $k^-$ plane, depending on the region of $x$. To the left
  (right) are the singularities above (below) the real $k^-$ axis. The
  diagrams show the new cut as one changes from one region of $x$ to
  the next, going from top to bottom on the left and from bottom to
  top on the right. We also give an axis in $x'$ to display the
  symmetry between the two partons.}
\end{figure}

Let us now discuss in turn the regions of $x$ in the interval $\xi-1 <
x < 1$. Details of the argument are given in the Appendix.
\newcounter{case}
\begin{list}{\Alph{case}.}{\usecounter{case}
    \setlength{\leftmargin}{\parindent}}
\item In the region $\xi\le x<1$, the $s$~singularities are in the
  upper half plane of $k^-$ whereas all other singularities are in the
  lower half plane. Wrapping the $k^-$ contour around the $s$~cuts
  gives
  \begin{equation}
    \label{disc-s}
    \int_{-\infty}^{+\infty} dk^- \, {\cal A} = \int dk^- \, {\rm
    disc}_s{\cal A}  \eqcm
  \end{equation}
  where ${\rm disc}_s{\cal A}$ is the discontinuity of $\cal A$ across
  the cuts in $s$. We use now the standard reasoning on the matrix
  element
  \begin{eqnarray}
    \label{quark-dist}
  \widehat{M} &=&
  \frac{1}{2 \pi}  \int dz^- \, e^{i x p^+ z^-}
  \left. \mat{p'}{\bar{\psi}(0) \gamma^+ \psi(z)}{p} \right|_{z =
    z^- \, v'}  \nonumber \\
  &=& \frac{1}{(2\pi)^4}  \int d^2 {\bf k}_T \, d k^-
  \left. \int d^4 z \, e^{i k \cdot z} \mat{p'}{\bar{\psi}(0)
      \gamma^+ \psi(z)}{p}  \right|_{k^+ =x p^+}
  \end{eqnarray}
  which does not involve time ordering, inserting a complete set of
  intermediate states between $\bar\psi(0)\gamma^+$ and $\psi(z)$ as
  illustrated in Fig.~\ref{fig:cuts} $(a)$. In line with our assumptions 
  on its analyticity properties we assume that one can apply the cutting
  rules to ${\cal A}$ as one would for Feynman diagrams, and obtain
  \begin{equation}
    \label{same-s}
    M = \frac{1}{(2\pi)^4}  \int d^2 {\bf k}_T \, d k^- \, {\rm
    disc}_s {\cal A} = \widehat{M}  \eqpt
  \end{equation}
  
  In this region of $x$ we can interpret $M$ as the amplitude for the
  proton emitting a quark with light cone fraction $x$ and reabsorbing
  it with light cone fraction $x'$.
\item In the region $\xi-1 < x \le 0$, the $u$~singularities are
  alone below the real $k^-$ axis and we can write
  \begin{equation}
    \label{disc-u}
  \int_{-\infty}^{+\infty} dk^- \, {\cal A} = \int dk^- {\rm disc}_u
  {\cal A}  \eqpt
  \end{equation}
  We now consider the scattering process
  \begin{equation}
    \label{other-scattering}
    p(p) + q(k') \to p(p') + q(k)  \eqcm
  \end{equation}
  whose amplitude is given by $- {\cal A}$, where the minus sign
  comes from interchanging the quark and antiquark field under the
  time ordering in (\ref{amplitude}). Defining
  \begin{equation}
    \label{antiquark-dist}
  \widehat{M}' = -
  \frac{1}{2 \pi}  \int dz^- \, e^{i x p^+ z^-}
  \left. \mat{p'}{\psi(z) \; \bar{\psi}(0) \gamma^+}{p} \right|_{z =
    z^- \, v'}  \eqcm
  \end{equation}
  and repeating our above argument we find
  \begin{equation}
    \label{same-u}
     M = \frac{1}{(2\pi)^4}  \int d^2 {\bf k}_T \, d k^- \, {\rm
    disc}_u {\cal A} = \widehat{M}'  \eqpt
  \end{equation}
  
  In the present region of $x$ we can interpret $M$ as corresponding
  to the emission by the proton of an antiquark with momentum fraction
  $-x'$ and its reabsorption with momentum fraction $-x$.
\item The intermediate region, $0<x<\xi$, is more involved since the
   singularities in $s$ and $k'^2$ lie in the upper half plane of
   $k^-$ while those in $u$ and $k^2$ lie in the lower half
   plane.\footnote{The particular singularity structure of $\cal A$ in
   this region of $x$ has also been remarked in \cite{RadLong,FSF}.}
   We notice that in this case $x>0$ and $x'<0$, i.e.\ in this
   situation one has a quark and an antiquark flowing out of the
   initial state. This configuration is specific of the off-diagonal
   matrix element, and in~\cite{RadComptMeson,RadLong} its
   reminiscence of the quark-antiquark distribution amplitude of a
   meson has been emphasised.
  
  If we choose to close the $k^-$ contour in the upper half plane we
  can write the integral over $k^-$ of $\cal A$ as a sum of terms due
  to the singularities in $s$ and in $k'^2$,
  \begin{equation}
    \label{disc-s-k}
    \int_{-\infty}^{+\infty} dk^- \, {\cal A} = \int dk^- \, 
    \Big( {\rm disc}_s  {\cal A} +
    \{ \mbox{terms from $k'^2$ singularities} \} \Big)  \eqcm
  \end{equation}
  where the singularities in $k'^2$ are cuts and a mass pole, cf.~the
  Appendix. Having chosen this representation we consider again
  $\widehat{M}$ and the scattering (\ref{scattering}), but now when we
  insert a complete set of states, the kinematics allow intermediate
  states that already contain the outgoing proton, $|X \rangle = |p'
  X' \rangle$. For those states there are diagrams which are
  disconnected at the right-hand side of the cut,
  cf.~Fig.~\ref{fig:cuts} $(b)$, and correspond to a cut in the
  variable $k'^2$. They give just the extra terms in (\ref{disc-s-k})
  and the cutting rules now allow us to write
  \begin{equation}
  \label{same-s-k} 
    M = \frac{1}{(2\pi)^4} \int d^2 {\bf k}_T \, d k^-
    \Big( {\rm disc}_s  {\cal A} +
    \{ \mbox{terms from $k'^2$ singularities} \} \Big) =
    \widehat{M} \eqpt 
  \end{equation}
\end{list}
In all three regions of $x$ there are of course two ways to pick up
singularities in the $k^-$ plane: those in the upper half will lead to
$\widehat{M}$ and those in the lower half to $\widehat{M}'$. The
situation is summarised in Fig.~\ref{fig:cuts}. In regions~A and B we
have a simple physical interpretation if we pick up the $s$ and
$u$~cut, respectively, as was already discussed for diagonal densities
in~\cite{Jaf}, whereas in region~C the choices $s$, $k'^2$ and $u$,
$k^2$ appear as symmetric to each other.

For all values of $x$ we then have
\begin{equation}
  \label{all-same}
  \rule[-2ex]{0pt}{6ex}     
  M = \widehat{M} = \widehat{M}'   \eqcm
\end{equation}
i.e.\ we can drop the time ordering in $M$ and write quark and
antiquark field in any order. Our argument shows that the origin of
this is the integration over $k^-$, which corresponds to fields being
evaluated at the same light cone time ($z^+=0$). We could in fact
apply it without integrating over ${\bf k}_T$ at all to obtain the
analogue of (\ref{all-same}) for ${\bf k}_T$ unintegrated parton
densities.

\vspace{\baselineskip}
\noindent
6. Our preceding discussion can be applied to off-diagonal gluon
distributions in an analogous way by considering the amplitude for the
scattering of an off-shell gluon on the proton. Note that one then
deals with products of gluon \emph{potentials} $A^\mu$, while the
gluon distributions are defined in terms of gluon \emph{field
strengths} $G^{\mu \nu}$. The passage from one to the other is however
simple in the $A^+ = 0$ gauge, since the leading twist distributions
only involve $G^{+ \nu}$ which in this gauge reduces to $\partial^+
A^\nu$. For a further discussion we refer to~\cite{RadLong}.

In the case of gluons there is no minus sign when one interchanges the
order of the fields in the amplitude (\ref{amplitude}) as there was in
point B.\ above, and correspondingly the analogue of $\widehat{M}'$ is
defined without a minus sign.

We also remark that there are symmetry relations for the gluon
distributions under the change $x \to -x' = \xi-x$ in their first
argument, so that it is enough to know them for $x \ge \xi /2$.  In
fact the diagrams in Fig.~\ref{fig:factorise} with gluon lines between
$S$ and $H$ remain the same if one interchanges the light cone
fractions $x$ and $-x'$, since $H$ denotes a sum of several Feynman
graphs.

\vspace{\baselineskip}
\noindent
7. As mentioned in the beginning, the diagrams of
Fig.~\ref{fig:factorise} $(b)$ for exclusive meson production contain
not only off-diagonal parton distributions as nonperturbative input,
but also the meson distribution amplitude, which is expressed in terms
of a matrix element between the vacuum and the meson state $|M
\rangle$. It is easy to show along the lines of argument developed so
far that the time ordering of quark operators in the distribution
amplitude can be omitted, just as in parton distributions, as we will
briefly outline.

\begin{figure}
  \epsfysize 4.2cm
  $$\epsfbox{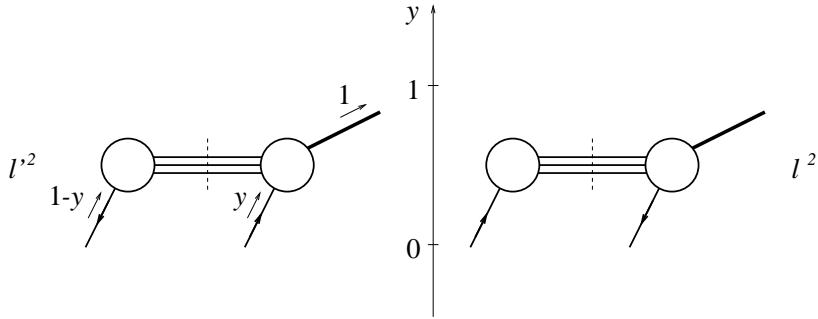}$$
  \caption{\label{fig:meson}The two possibilities to pick up
    singularities in the $l^+$ plane for a meson distribution
    amplitude in the region $0 < y < 1$.}
\end{figure}

The distribution amplitude is defined in terms of a two-point function
\begin{equation}
\frac{1}{2 \pi}  \int dz^+ \,
  e^{-i y q'^- z^+} \left. \mat{M}{T \, \bar{\psi}(z) \gamma^-
  \psi(0)}{0} \right|_{z = z^+ \, v}
\end{equation}
or a corresponding one with a different Dirac matrix; in order to
suppress the path ordered exponential between the quark fields one now
needs the $A^- = 0$ gauge. In analogy to (\ref{rewrite}) this
two-point function can be rewritten as an integral over ${\bf l}_T$
and $l^+$ at fixed $l^- = y q'^-$ of the amplitude for the transition
\begin{equation}
q(l) + \bar{q}(l') \to M(q')
\end{equation}
from the valence quarks to the meson. This amplitude depends on the
invariants $l^2$ and $l'^2$. Writing
\begin{eqnarray}
l^+ &=& \frac{l^2 + {\bf l}_T^2}{2 q'^- y} \eqcm \nonumber \\
l^+ &=& \frac{l'^2 + ({\bf l}_T - {\bf \Delta}_T)^2}{2 q'^- (y -1)} 
        + q'^+ 
\end{eqnarray}
and closing the contour of the $l^+$ integration in the complex plane
one immediately obtains that the distribution amplitude vanishes
outside the region $0 < y < 1$. There one can express it in terms
of singularities in either $l'^2$ or $l^2$, obtaining matrix elements
of $\bar{\psi}(z) \gamma^- \psi(0)$ or $- \psi(0) \, \bar{\psi}(z)
\gamma^-$, respectively. The two possibilities are shown in
Fig.~\ref{fig:meson}.

Comparing with the present situation we see again the hybrid nature of
off-diagonal parton distributions in the region $0<x<\xi$\,: the cuts
in $s$ or $u$ are as in ordinary, diagonal parton distributions,
whereas the singularities in the parton virtualities $l^2$ or $l'^2$
are reminiscent of distribution amplitudes.

\vspace{\baselineskip}
\noindent
8. The fact that parton distributions and meson distribution
amplitudes can be expressed in terms of \emph{cut} amplitudes is
important if one wants to show that the $s$-channel discontinuities of
the $\gamma^\ast p$ amplitudes shown in Fig.~\ref{fig:factorise} are
obtained from appropriate cuts of the hard blob $H$, the soft
quantities being ``already cut''. Already of interest in the case of
inclusive deep inelastic scattering this is essential for processes
involving off-diagonal parton distributions, for instance when
establishing a dispersion relation~\cite{Col}. We shall not go into
details here but wish to make some observations in connection with the
results we have obtained so far.

There are different ways to cut the diagrams of
Fig.~\ref{fig:factorise} in the overall $s$-channel, i.e.\ in the
variable $(p + q)^2$. If the hard blob $H$ is cut in $(q + k)^2$,
which requires $\xn \le x \le 1$ to ensure positive energy across the
cut, one can cut the soft blob $S$ in $s = (p - k)^2$; if $x \le \xi$
one can also cut $S$ in $k'^2$ or cut the parton line $k'$ itself in
the diagrams. The two-point function $M$ can be written in terms of
just these cuts. As discussed in the appendix it includes in
particular the term that comes from cutting the parton line $k'$ and
leads to a mass pole in the amplitude $\cal A$ of (\ref{amplitude});
hence this mass pole does not appear as an extra term in the
expression of the discontinuity of the $\gamma^\ast p$ amplitude.

$H$ can also be cut in $(q - k')^2$, then one must have $\xi-1 \le x
\le \xi-\xn$. This comes with cuts of $S$ in $u = (p + k')^2$, and if
$x \ge 0$ also with cuts in $k^2$ as well as the pole from cutting the
parton line $k$. In this situation one will use the representation of
$M$ in terms of singularities in $u$ and $k^2$. It is interesting to
note that if there is a region $\xn \le x \le \xi-\xn$, e.g.\ in
photoproduction of a heavy meson or a $Z$, one can have cuts of $H$ in
both $(q + k)^2$ and $(q - k')^2$ at the same value of $x$ and needs
both representations of $M$ at the same time.

In the case of meson production (Fig.~\ref{fig:factorise} $(b)$) one
can also cut through the blob representing the meson distribution
function, with one valence quark at either side of the cut, or one can
cut through one of the valence quark lines.\footnote{Note that such
cuts already appear if one takes the hard blob $H$ at Born level, cf.\
e.g.\ the diagrams in Fig.~2 of \cite{Col}.} In this case one will use
that the meson distribution amplitude can be written in terms of
singularities in one of the quark virtualities $l^2$ and $l'^2$ as
discussed in sec.~7.

\vspace{\baselineskip}
\noindent{\bf Acknowledgments} \\* 
We gratefully acknowledge discussions with Jochen Bartels, Martin
Beneke, John Collins, Peter Landshoff, Piet Mulders, Otto Nachtmann
and Bernard Pire. Special thanks are due to O. Nachtmann for reading
the manuscript. M. D. would like to thank the NIKHEF Theory Group for
its hospitality.

T. G. was carrying out his work as part of a training Project of the
European Community under Contract No.~ERBFMBI--CT95--0411. This work
has been partially funded through the European TMR Contract
No.~FMRX--CT96--0008: Hadronic Physics with High Energy
Electromagnetic Probes. CPT is Unit\'e Propre du CNRS and SUBA\-TECH
is Unit\'e Mixte de l'Universit\'e de Nantes, de l'Ecole des Mines de
Nantes et du CNRS.

\vspace{\baselineskip}
\noindent{\bf Appendix} \\* In this appendix we study in some detail
the identity of the matrix elements $M$ and $\widehat{M}$ introduced
respectively in~(\ref{rewrite}) and~(\ref{quark-dist}). As we already
mentioned the ${\bf k}_T$ integration has no relevance in the
derivation so that we perform our reasoning on the corresponding ${\bf
k}_T$ unintegrated quantity. We focus on the most complicated region
$0<x<\xi$ where as discussed in point C.\ of sec.~5 the integration
over $k^-$ inevitably picks up singularities in more than one
invariant and where the cutting rules have to be applied with some
care. In order to deal with complete but readable expressions we
ignore here complexities due to the spin of both partons and hadrons
and introduce
\begin{eqnarray}  \label{scalar-matrix-elements}
{\cal M} &=& \int_{-\infty}^{+\infty} dk^- \int\! d^4z\, e^{ik\cdot z}\,
\langle p'|T\phi^{\dagger}(0)\phi(z)|p\rangle \eqcm \\ \nonumber
\widehat{{\cal M}} &=& \int_{-\infty}^{+\infty} dk^- \int\! d^4z\, 
e^{ik\cdot z}\, \langle p'|\phi^{\dagger}(0)\phi(z)|p\rangle \eqcm
\end{eqnarray}
where $\phi$ is a charged scalar ``quark'' field and $|p\rangle$ a
scalar ``proton'' state.

In the framework we are using we invoke perturbation theory for the
purpose of analytic properties and of applying the cutting rules. To
be consistent we must treat quarks as free particles with a mass
$m_q$. The scalar analogue of the amplitude $\cal A$ in
(\ref{amplitude}) then has cuts and a mass pole in each of $k^2$ and
$k'^2$. To apply the cutting rules we will isolate the poles from the
rest of the singularity structure, and also keep track of possible
disconnected terms of the matrix elements that appear in our argument.
To this end we use the framework of the reduction formula \cite{LSZ}
(cf.\ also \cite{ItzZub}) and introduce an interpolating field
$\mathit{\Phi}$ for the proton. For $p \neq p'$ the matrix element
$\langle p' |T\phi^{\dagger}(0)\phi(z) |p\rangle$ does not have a
disconnected part, and in the case of diagonal parton densities one
explicitly only takes its connected part. One thus has
\begin{equation} \label{pqtopq}
\int\! d^4z\, e^{ik\cdot z}\,
\langle p'|T\phi^{\dagger}(0)\phi(z)|p\rangle =
{i\sqrt{Z_q}\over k'^2-m_q^2+i\varepsilon}\,
{i\sqrt{Z_q}\over k^2-m_q^2+i\varepsilon}\, 
i \tilde{G}(p\bar{q}\to p'\bar{q}')
\end{equation}
with
\begin{eqnarray} \label{def-G-tilde}
\lefteqn{ (2\pi)^4 \delta^{(4)}(p-k-p'+k') \cdot i \tilde{G} =
\frac{1}{Z_q}\, \frac{1}{Z_p}\,
\int d^4z\, d^4z'\, d^4y\, d^4y'\, e^{ik\cdot z - ik'\cdot z' -ip\cdot
y +ip'\cdot y'} \cdot}  \nonumber \\
& & (\Box_{z'} + m_q^2)\, (\Box_{z} + m_q^2)\,
(\Box_{y'} + m_p^2)\, (\Box_{y} + m_p^2) \cdot
\langle 0 |T \phi^{\dagger}(z')\phi(z) 
\mathit{\Phi}(y') \mathit{\Phi}^{\dagger}(y)| 0\rangle  \eqcm
\end{eqnarray}
where $Z_q$ and $Z_p$ are the wave function normalisation constants of
the quark and the proton, and where four-momenta are labelled as
in~(\ref{scattering}). In the limit where both parton legs go on shell
$\tilde{G}$ is a $\cal T$-matrix element: depending on the signs of
$k^0$ and $k'^0$ the partons are in the initial or final state. We
emphasise that only the parton mass poles in $k^2$ or $k'^2$ have been
isolated from $\tilde{G}$ which thus still contains the branch cuts in
these variables, unlike truncated Green functions where the full
propagators including these cuts are split off.

To obtain ${\cal M}$ we integrate the right hand side of
eq.~(\ref{pqtopq}) over $k^-$ and close the contour in the upper half
plane, assuming that convergence is fast enough at infinity. In the
region $0<x<\xi$ the only singularities in the upper half $k^-$ plane
are due to the pole $1 /(k'^2-m_q^2+i\varepsilon)$ and the cuts in $s$
and $k'^2$ of $\tilde{G}$, according to our hypothesis on the
singularity structure. We get
\begin{eqnarray} \label{expr-M}
{\cal M} &=& \int_{-\infty}^{+\infty} dk^-\, 
2\pi\delta(k'^2-m_q^2) \sqrt{Z_q}\,
{i\sqrt{Z_q}\over k^2-m_q^2 + i\varepsilon}\, 
i\tilde{G}(p\bar{q}\to p'\bar{q}')  \nonumber \\
&+& \int_{-\infty}^{+\infty} dk^- 
{i\sqrt{Z_q}\over k'^2-m_q^2 - i\varepsilon}\,
{i\sqrt{Z_q}\over k^2-m_q^2 + i\varepsilon}\,
i \Big[{\rm disc}_s\tilde{G}+{\rm disc}_{k'^2}\tilde{G}\Big] \eqpt
\end{eqnarray}
Note that the sign of $i\varepsilon$ in the $k'^2$ pole has changed
from (\ref{pqtopq}) to (\ref{expr-M}) as a consequence of separating
the contributions of the pole and the cuts, with the pole lying below
the cuts in the $k^-$ plane.

Using the cutting rules \cite{Cut} (cf.\ also \cite{ItzZub}) the above
discontinuities can be expressed as
\begin{eqnarray}
-i\, {\rm disc}_s\tilde{G}(p\bar{q}\to p'\bar{q}') &=&
\sum_X (2\pi)^4\delta^{(4)}(p_X + k' - p')\,
\tilde{G}(p\bar{q}\to X)\,\tilde{G}^*(p'\bar{q}'\to X) 
\eqcm \label{cutting-rules-s} \\
-i\, {\rm disc}_{k'^2}\tilde{G}(p\bar{q}\to p'\bar{q}') &=&
\!\!\!\sum_{X'\ne \bar{q}''} (2\pi)^4 \delta^{(4)}(p_{X'} + k') \,
\tilde{G}(p\bar{q}\to p'X')\,\tilde{G}^*(\bar{q}'\to X') \eqcm
 \label{cutting-rules-k}
\end{eqnarray}
where $p_X$ and $p_{X'}$ respectively denote the four-momenta of the
cut states $X$ and $X'$. The functions $\tilde{G}$ are defined in
analogy to (\ref{def-G-tilde}), i.e.\ as Fourier transformed Green
functions with the poles in the external legs removed. They reduce
again to $\cal T$-matrix elements in the on-shell limit of the
external antiquark legs. The $k'^2$ cut (\ref{cutting-rules-k}) has no
contribution from a single antiquark state, as expressed in the
restriction $X'\ne \bar{q}''$, where $\bar{q}''$ has momentum $-
k''$\,: if $\bar{q}'$ is off-shell such a state is excluded by
momentum conservation and if it is on-shell then the amplitude
$\tilde{G}$ is zero for the non-interacting transition
$\bar{q}'\to\bar{q}''$.

Let us now turn to the study of $\widehat{{\cal M}}$ when
$0<x<\xi$. Inserting a complete set of ``out'' states\footnote{So far
we only had matrix elements between one-particle states or the vacuum,
which are equal in the ``in'' and ``out'' representations, here we
need to make the distinction.} we obtain
\begin{equation}  \label{complete-set}
\widehat{{\cal M}}=\int_{-\infty}^{+\infty} dk^- \sum_X 
(2\pi)^4\delta^{(4)}(p_X+k'-p') \;
\in\langle p'|\phi^{\dagger}(0)|X\rangle\out \cdot
\out\langle X|\phi(0)|p\rangle\in  \eqpt
\end{equation}
Due to momentum conservation $|X\rangle$ cannot contain the initial
proton state $|p\rangle$ in the region $0<x<\xi$ we are
considering. The second matrix element in (\ref{complete-set}) is then
connected and can be written in terms of $\tilde{G}$ using again the
reduction formula:
\begin{equation}\label{pqtoX}
\out \langle X|\phi(0)|p\rangle\in =
{i\sqrt{Z_q} \over k^2-m_q^2+i\varepsilon}\,  
i\tilde{G}(p\bar{q}\to X)  \eqpt
\end{equation}
There are however states that contain the final proton state
$|p'\rangle$ and lead to disconnected parts of $\in\langle
p'|\phi^{\dagger}(0)|X\rangle\out$.  For states
$|X\rangle=|p''X'\rangle$ with a proton $|p''\rangle$ of momentum
$p''$ we have
\begin{eqnarray} \label{pprime}
\out\langle p''X'|\phi(0)|p' \rangle\in &=&
{i\sqrt{Z_q} \over k'^2-m_q^2+i\varepsilon}\,  
i\tilde{G}(p'\bar{q}'\to p''X')  \nonumber \\ 
&+& (2\pi)^3 \,2{p'}^0\, \delta^{(3)}({\bf p}'-{\bf p}'') \,
{i\sqrt{Z_q}\over k'^2-m_q^2 +i\varepsilon}
i\tilde{G}(\bar{q}'\to X')  \eqpt
\end{eqnarray}
If $X'$ is a single antiquark state $\bar{q}''$, which due to momentum
conservation requires $\bar{q}'$ to be on shell, the disconnected term
at the r.h.s.\ of (\ref{pprime}) simply reads
\begin{equation}
(2\pi)^3 \,2{p'}^0\, \delta^{(3)}({\bf p}'-{\bf p}'')\,
  \out\langle \bar{q}''|\phi(0)|0\rangle\in =
(2\pi)^3 \,2{p'}^0\, \delta^{(3)}({\bf p}'-{\bf p}'')\, \sqrt{Z_q}
  \eqpt
\end{equation}
For states $|X\rangle$ that do not contain a proton $|p''\rangle$ the
analogue of (\ref{pqtoX}) holds. Together with the connected term in
(\ref{pprime}) these states give the $s$-channel cut in ${\cal M}$,
whereas the disconnected terms give the cuts and pole in $k'^2$. Using
\begin{equation}
\sum_{X'=\bar{q}''} (2\pi)^4\delta^{(4)}(p_{X'}+k') 
= 2\pi\delta(k'^2-m_q^2)
\end{equation}
for the pole term we obtain \pagebreak
\begin{eqnarray}  \label{expr-M-tilde}
\lefteqn{ \widehat{{\cal M}} = \int_{-\infty}^{+\infty} dk^- \Biggl\{
2\pi\delta(k'^2-m_q^2) \sqrt{Z_q} \, 
{i\sqrt{Z_q}\over k^2-m_q^2+i\varepsilon}\,
i\tilde{G}(p\bar{q}\to p'\bar{q}') }  \nonumber \\
&+& \sum_{X'\ne \bar{q}''}
(2\pi)^4\delta^{(4)}(p_X'+k')
{i\sqrt{Z_q}\over k^2-m_q^2 +i\varepsilon}\,
{-i\sqrt{Z_q}\over k'^2-m_q^2 -i\varepsilon}\,
\tilde{G}(p\bar{q}\to p'X')\, \tilde{G}^*(\bar{q}'\to X')  \nonumber \\
&+& \sum_X (2\pi)^4\delta^{(4)}(p_X+k'-p')
{i\sqrt{Z_q}\over k^2-m_q^2+i\varepsilon}\,
{-i\sqrt{Z_q}\over k'^2-m_q^2-i\varepsilon}\,
\tilde{G}(p\bar{q}\to X)\, \tilde{G}^*(p'\bar{q}'\to X) \Biggr\}  \eqpt
\nonumber \\
\end{eqnarray}
With (\ref{expr-M}) and the cutting rules (\ref{cutting-rules-s}),
(\ref{cutting-rules-k}) we finally have $\widehat{{\cal M}} = {\cal
M}$, and thus $\widehat{M}=M$.\footnote{In the region $\xi \le x < 1 $
discussed in point A.\ of sec.~5 some attention is required to see
that the signs of $i\epsilon$ in the $k^2$ and $k'^2$ poles are
compatible between $\widehat{{\cal M}}$ and ${\cal M}$. Assuming that
${\rm disc}_s\tilde{G}$ vanishes for $s$ below $m_p^2$ due to its
threshold one finds however that these signs can actually be chosen
arbitrarily.}

\end{document}